\documentclass[twocolumn,showpacs,preprintnumbers,amsmath,amssymb]{revtex4}
\usepackage{amsmath}
\usepackage[usenames]{color}
\usepackage[dvips]{graphics}

\begin{document}
\draft
\title{The Phase Transition of Square Lattice Antiferromagnets at Finite Temperature}
\author{Huai-Yu Wang}
\date{\today }
\draft

\address{Department of Physics, Tsinghua University, Beijing 100084, China}

\begin{abstract}
The magnetic properties of the two-dimensional $J_{1}-J_{2}$ model
with both exchanges $J_{1}$ and $J_{2}$ being antiferromagnetic and
a single-ion anisotropy at nonzero temperature are investigated. As
$J_{2}/J_{1}<1/2$ ($>1/2$), only the N\'{e}el (collinear) state
exists. When $J_{2}/J_{1}=1/2$, both the N\'{e}el and collinear
states can exist and have the same N\'{e}el temperature. The
calculated free energies show that there can occur a phase
transition between the two states below the N\'{e}el point when the
single-ion anisotropy is strong enough. It is a first-order
transition at nonzero temperature. It is possible that the doping in
real materials can modify the ratio of $J_{2}/J_{1}$ to reach $1/2$
so as to implement the phase transition.
\end{abstract}


\maketitle


\newpage

A square lattice antiferromanget can be described by the well-known
two-dimensional (2D) $J_{1}-J_{2}$ model. In this model both the
nearest neighbor (nn) and next nearest neighbor (nnn) exchanges are
antiferromagnetic (AF), so that it was believed a frustrated system.
This model could be used to describe the structures in real
materials. It was firstly related to the copper oxide monolayers in
the Cu-based high temperature oxide superconductors[1], and then to
magnetic planes in some other materials[2-6]. The most
representative structure that could be well described by this model
might be the Fe monolayers in the Fe-based superconductors
La-O-Fe-As[7-13] and BaFe$_{2}$As$_{2}$[14]. Because of its
importance, the $J_{1}-J_{2}$ model has been carefully studied by
various methods. However, the study has been mainly focused on its
properties, especially its possible phase transition at 0K[15-17].
Investigations concerning nonzero temperature[18-24] have been
comparatively much fewer, although the real materials are at finite
temperature. Despite the already given physical results of the
system by these investigations, there may be some interesting
features still hidden at finite temperature.

Here we study the $J_{1}-J_{2}$ model as a representative of such a
Fe plane, focusing our attention on the properties at nonzero
temperature. The physical quantities of the quantum model at finite
temperature are calculated. A remarkable result is that when
$J_{2}/J_{1}=1/2$, there may occur a phase transformation below
N\'{e}el point $T_{N}$.

The AF Hamiltonian of a square lattice is
$H=\frac{1}{2}\sum\limits_{i,j}J_{ij}\textbf{\emph{S}}_{i}\cdot
\textbf{\emph{S}}_{j}-D\sum\limits_{i} (S^{z}_{i})^{2}.$ The first
term is Heisenberg exchanges. Only the nn and nnn exchanges $J_{1}$
and $J_{2}$ are considered, both being positive. The second term
presents a single-ion anisotropy. If a two-dimensional AFM system
has no any anisotropy, there will be no spontaneous sublattice
magnetizations in it[24,25]. It was indeed possible for the
single-ion anisotropy to appear in real materials[26]. We term the
exchanges $J_{1}$ and $J_{2}$ and anisotropy $D$ as Hamiltonian
parameters. We let Boltzman constant $k_{B}=1$ so that all the
quantities, including Hamiltonian parameters, temperature $T$, and
sublattice magnetization $\langle S^{z}\rangle$, become
dimensionless. $\langle S^{z}\rangle$  is the assembly
thermostatistical average of spin operator $S^{z}$. We fix $J_{1}=1$
and change $J_{2}$ value in computation. $D$ is assumed to be two or
three orders of magnitude less than $J_{1}$. In the real La-O-Fe-As
materials the spin quantum number might be larger than 1/2[27].
Therefore, the cases of some of the lowest spin quantum numbers are
considered.

\begin{figure}
\centering \scalebox{0.38}[0.40]{\includegraphics{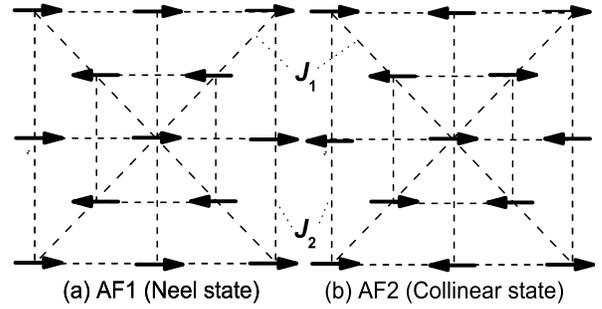}}
\caption{ (a) AF1 and (b) AF2 configurations.}
\end{figure}

It was proposed that there might be four possible spin
configurations[28,29], among which the two named as AF1 and AF2[30]
had lower energies. They are depicted in Fig. 1, and were called as
N\'{e}el state and collinear state, respectively. In either of the
configurations, the lattice is divided into two sublattices. The
spins within each sublattice are parallel to each other, and the
spins of the two sublattices antiparallel to each other. The spin
averages of the two sublattices are denoted as $\langle
S^{z}_{1}\rangle$ and $\langle S^{z}_{2}\rangle$, respectively.
Since there is no external field, $\langle S^{z}_{1}\rangle=-\langle
S^{z}_{2}\rangle=\langle S^{z}\rangle$. We calculate the stable
configurations by the many-body Green's function method under random
phase approximation[31]. According to our calculation results, when
$J_{2}/J_{1}<1/2$, the stable state is AF1 configuration where the
nn spins are antiparallel to each other, showing that the nn
exchange is dominant. While for $J_{2}/J_{1}>1/2$, the stable state
is AF2 where the nnn spins are antiparallel to each other. This
conclusion holds at any temperature, for any $S$ and nonzero $D$
values. We also tested other ordered states including the two
suggested in Refs. [28,29] and none of them were stable at nonzero
temperature.

\begin{figure}
\centering \scalebox{0.36}[0.34]{\includegraphics{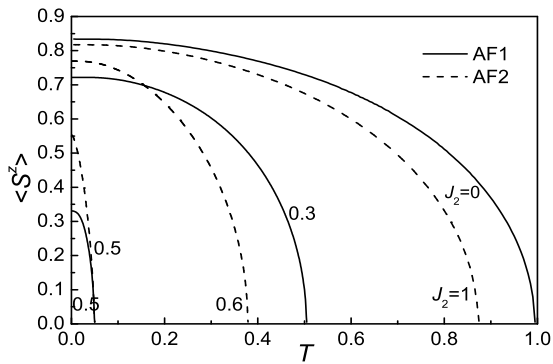}}
\caption{ $\langle S^{z}\rangle$ vs. $T$ curves for $S$=1 and
$D$=0.01. The numbers labeling the curves are the $J_{2}$ values.
When $J_{2}<0.5 (>0.5)$, the state is AF1 (AF2). When $J_{2}=0.5$,
both states exist.}
\end{figure}

In Fig. 2 we plot the curves of spin average $\langle S^{z}\rangle$
vs. $T$ at various $J_{2}$ values. In this and following figures, we
always use the solid and dashed lines to represent the results of
AF1 and AF2 configurations, respectively. The temperature at which
$\langle S^{z}\rangle$ becomes zero is N\'{e}el point, denoted as
$T_{N}$. Because we merely research the case of nonzero temperature,
when we mention zero temperature, we actually mean the temperature
very close to zero, which is denoted by $0^{+}$. To describe the
dependence of the curves on $J_{2}$ value, we concentrate our
attention to two physical quantities: N\'{e}el temperature $T_{N}$
and the spin average $\langle S^{z}\rangle$ at $0^{+}$K, hereafter
denoted as $\langle S^{z}(0^{+})\rangle$. As $J_{2}$ increases from
zero to 0.5, both N\'{e}el temperature $T_{N}$ and $\langle
S^{z}(0^{+})\rangle$ decreases. At $J_{2}=0$, it is an ordinary nn
AF exchange system, and there is no competition to cause
frustration. As $J_{2}$ increases from 0, the competition between
$J_{2}$ and $J_{1}$ emerges and becomes stronger. This results in
the drop of both $T_{N}$ and $\langle S^{z}(0^{+})\rangle$.

When $J_{2}=0.5$, the competition between $J_{1}$ and $J_{2}$ is the
strongest. As $J_{2}$ rises from 0.5, the role of $J_{2}$ becomes
more important and the competition becomes comparatively weaker. As
a consequence, both $T_{N}$ and $\langle S^{z}(0^{+})\rangle$
increase.

Figure 3 plots the results of $T_{N}$ as a function of $J_{2}$ at
different $S$ and $D$ values. This figure is in fact a phase diagram
that contains three phases: AF1, AF2 and paramagnetic (P) phases. A
solid line is the border line between phases AF1 and P, and a dashed
one is the border between AF2 and P. As $J_{2}/J_{1}$ approaches 1/2
from either side, the competition between $J_{1}$ and $J_{2}$
becomes stronger and makes $T_{N}$ lower. At $J_{2}/J_{1}=1/2$,
$T_{N}$ is the lowest. Calculations show that when $J_{2}>1$,
$T_{N}$ is linearly proportional to $J_{2}$.

\begin{figure}
\centering \scalebox{0.32}[0.28]{\includegraphics{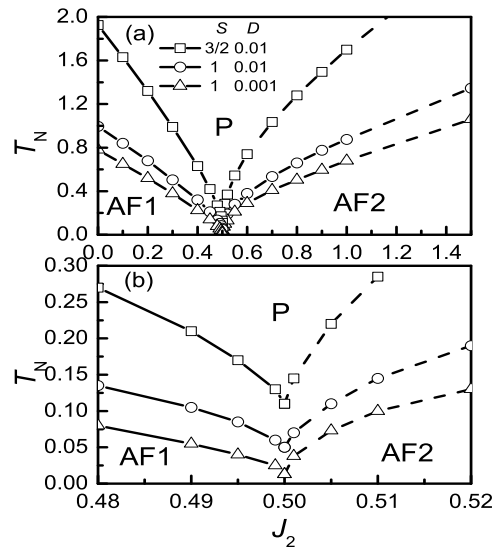}}
\caption{(a) N\'{e}el temperature vs. $J_{2}$ value. (b) Enlargement
of the region around $J_{2}=0.5$. The lines are to guide eyes.}
\end{figure}

\begin{figure}
\centering \scalebox{0.32}[0.28]{\includegraphics{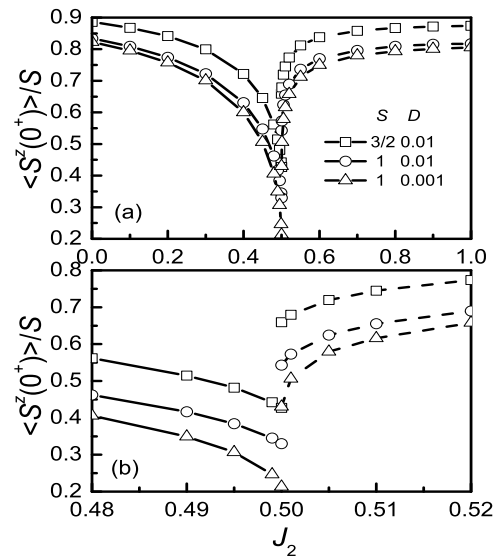}}
\caption{(a) $\langle S^{z}(0^{+})\rangle$ vs. $J_{2}$ value. (b)
Enlargement of the region around $J_{2}=0.5$. The lines are to guide
eyes.}
\end{figure}

Figure 4 plots curves of $\langle S^{z}(0^{+})\rangle$ vs. $J_{2}$
value. As temperature approaches to $0^{+}$K, the thermodynamic
fluctuation goes to zero but quantum fluctuation still exists in an
AF system. A smaller $\langle S^{z}(0^{+})\rangle$ value represents
a stronger quantum fluctuation. As $J_{2}/J_{1}$ approaches 1/2 from
either side, the competition between $J_{1}$ and $J_{2}$ lowers
$\langle S^{z}(0^{+})\rangle$ value.

As $J_{2}>1$, the curves of $\langle S^{z}(0^{+})\rangle$ vs.
$J_{2}$ value are almost flat. This means that when $J_{2}$ is
sufficiently large, it is predominant compared to $J_{1}$ value, so
that the quantum fluctuation at $0^{+}$K caused by the competition
between $J_{1}$ and $J_{2}$ is almost unchanged with the variation
of $J_{2}$ value. Nevertheless, no matter how large the $J_{2}$
value is, $\langle S^{z}\rangle$ of AF2 is always smaller than that
of AF1 when $J_{2}$=0. This is because when $J_{2}>0$, there is
always a competition between $J_{1}$ and $J_{2}$. While in the case
of $J_{2}$=0, there is no such a competition, and only the nn AF
exchange plays a role.

The effect of the anisotropy $D$ value and spin quantum number $S$
is embodied in Figs. 3 and 4. As $D$ or $S$ value is smaller,
$\langle S^{z}\rangle$ value at any temperature is smaller, and
N\'{e}el point $T_{N}$ is as well.

Now let us discuss the case of $J_{2}/J_{1}=1/2$. At first thought,
in this case the competition between $J_{1}$ and $J_{2}$ are
strongest so that the AF configurations may be totally frustrated
and either AF configuration is difficult to hold. Indeed,
researches[24] showed that as long as there was no single-ion
anisotropy, both configurations could not exist at finite
temperature when $J_{2}/J_{1}=1/2$. However, our calculation shows
that for any nonzero $D$ value, both AF1 and AF2 states can exist as
displayed by Figs. 2 to 4. In Fig. 2, the solid and dashed lines
marked by '0.5' show that both the states can exist and have the
same $T_{N}$ point under the parameter $J_{2}/J_{1}=1/2$. In Fig. 3,
it is shown that both AF1 and AF2 reach the same N\'{e}el
temperature when $J_{2}/J_{1}$ reaches 1/2 for various $S$ and $D$
values. Figure 4(b) reveals that both AF1 and AF2 have nonzero
$\langle S^{z}(0^{+})\rangle$ as $J_{2}/J_{1}$ reaches 1/2.

Since as $J_{2}/J_{1}$ goes to 1/2 from either side, $T_{N}$ reaches
the same value, see Fig. 3(b), it is understood that the N\'{e}el
point is uniquely determined by the Hamiltonian parameters, although
there may be more than one state.

Figure 4(b) shows that when $J_{2}/J_{1}=1/2$, $\langle
S^{z}(0^{+})\rangle$ values of the two states are not the same, and
that of AF2 is higher. This can be explained from the pictures in
Fig. 1. In AF2 configuration, every spin is parallel to a half of
its nn spins and antiparallel to another half, respectively, while
in AF1, every spin is antiparallel to all of its nn spins. Thus the
quantum fluctuation of AF2 at $0^{+}$K is stronger than AF1.

\begin{figure}
\centering \scalebox{0.38}[0.366]{\includegraphics{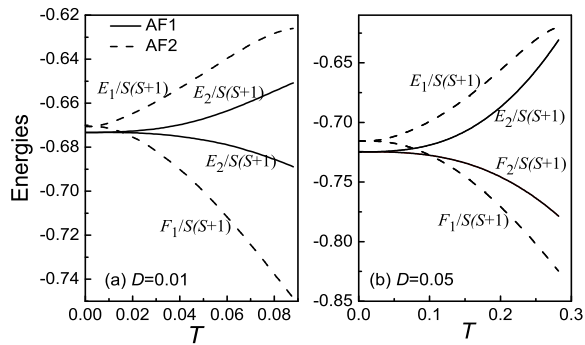}}
\caption{The internal energies $E(T)$ (ascending curves) and free
energies $F(T)$ (descending curves) at $J_{2}=0.5$ and $S=3/2$. (a)
$D= 0.01$ and (b) $D= 0.05$. }
\end{figure}

Since both configurations can exist, one may ask which one is
stabler. At fixed temperature and volume, the state with lower free
energy is stabler. The free energy can be evaluated numerically by
means of the internal energy via $
F(T)=E(0^{+})-T\int_{0^{+}}^{T}\frac{E(T')-E(0^{+})}{T'^{2}}dT' $.
Before calculating the free energy, one has to compute the internal
energy which is defined as the thermostatistical average of
Hamiltonian, $E=\langle H\rangle /N$, where $N$ is the site number
in the 2-D plane. The correlation functions
$\frac{1}{2}\sum\limits_{i,j}J_{ij}S^{+}_{i} S^{-}_{j}$ involved in
the energy are carefully calculated by use of the spectral
theorem[32]. Figure 5 plots $E(T)$ for $S=3/2$ and $D=0.01$ and
0.05. $E(T)$ increases with temperature monotonically as it should
be. In Fig. 5, $E_{2}(T)>E_{1}(T)$, but the internal energy cannot
be used to determine which stable is stabler, since the entropies of
the two states are different. The corresponding free energies are
plotted in Fig. 5. $F(T)$ decreases monotonically with temperature.
It is seen that $F_{1}(0^{+})<F_{2}(0^{+})$, which means that near
zero temperature AF1 configuration is stabler. However, $F_{2}(T)$
drops faster than $F_{1}(T)$ and the two curves cross at a
temperature, which means that above this temperature AF2 is stabler.
Thus it is concluded that below $T_{N}$, an AF1-AF2 phase
transformation may occur. By comparison of Figs. 5(a) and (b), it is
seen that as the anisotropy $D$ is raised from 0.01 to 0.05, both
$F_{1}(0^{+})$ and $F_{2}(0^{+})$ lower, the former decreasing more,
and thus the AF1-AF2 phase transformation point rises.

\begin{figure}
\centering \scalebox{0.40}[0.36]{\includegraphics{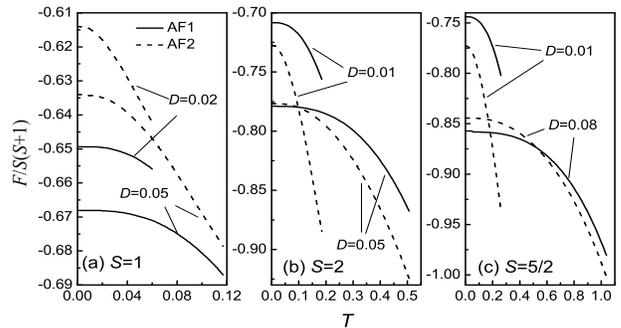}}
\caption{The free energies at $J_{2}=0.5$ and several $S$ and $D$
values. (a) $S=1$, (b) $S=2$ and (c) $S=5/2$. }
\end{figure}

In Fig. 6, we plot the free energy curves for $S=1, 2, 5/2$, each
with two $D$ values. The common features in this figure and Fig. 5
are that $F_{2}(T)$ always decreases faster than $F_{1}(T)$, and as
$D$ rises, both $F_{1}(0^{+})$ and $F_{2}(0^{+})$ drop, the former
dropping more. In Fig. 6(a), where $S=1$, for either $D$ value,
$F_{1}(0^{+})$ is much lower than $F_{2}(0^{+})$, and up to $T_{N}$,
the two curves do not cross. Therefore, in this case, there is no
phase transformation below N\'{e}el point. As for $S=3/2$, the
AF1-AF2 transformation may occur, as having been revealed by Fig. 5.
In Fig. 6(b), as $D=0.01$, $F_{1}(0^{+})$ is higher than
$F_{2}(0^{+})$, and up to $T_{N}$, the two curves do not cross,
indicating AF2 being always stabler and lack of the phase
transformation. While when $D$ is raised to 0.05, $F_{1}(0^{+})$
drops to such a position that $F_{1}(0^{+})< F_{2}(0^{+})$, and the
two curves $F_{1}(T)$ and $F_{2}(T)$ cross below $T_{N}$. Therefore,
a phase transformation may occur. The analysis of Fig. 6(c) is
similar to that of Fig. 6(b).

It is deduced from Figs. 5 and 6 that the condition for the AF1-AF2
phase transformation to occur is that the $D$ value should be large
enough so that $F_{1}(0^{+})< F_{2}(0^{+})$. Otherwise,
$F_{1}(0^{+})> F_{2}(0^{+})$ and there is no phase transformation,
because $F_{2}(T)$ always decrease with temperature faster than
$F_{1}(T)$. In the case of $S=1$, when the $D$ value continues to
increase, then both the solid and dashed lines lower, and if $D$ is
strong enough, it is expected that the two line will cross and the
phase transition will occur.

It should be noted that both AF1 and AF2 are stable states but with
different energies, so that it is a first-order transition. There is
certainly an energy barrier between the two states. Unlike a
classical system, the energy barrier in the present quantum system
is difficult to reckon  since it involves non-equilibrium states.

A question arises that how to actualized the AF1-AF2 phase
transformation. For instance, in the case of Fig. 5(b), suppose that
the system is initially under room temperature. When temperature is
decreased to below N\'{e}el point, the state becomes AF2. As
temperature reaches the AF1-AF2 transformation point, how the AF2
configuration can overcomes the barrier to get to AF1 configuration?
We suggest that applying a strong impulsive magnetic field along the
direction that is perpendicular to the spin direction can make the
system to reach the stabler state.

Finally, we would like to briefly discuss the possibility of
adjusting the ratio $J_{2}/J_{1}$ to become 1/2 in real materials.
The results from the band structure calculation of LaFeAsO were that
$J_{2}/J_{1}>1/2$[27,29,30,33], thus the Fe planes in this crystal
were in AF2 state. In this kind of materials both the nn and nnn
exchanges between Fe atoms were mediated by As atoms[27]. Because
the Fe 2-D plane was sandwiched by As atom monolayers, As atoms
played a key bridge role in the indirect exchanges. Since in AFeAsO
(A=La, Ce, Pr, Nd, Sm)[10,34] and BFe2As2 (B=Ca, Sr, Ba)[35] the
Fe-As sandwich structures were the same, the Fe planes in all these
crystals were in AF2 states. While LaFePO, which had the same
crystal structure as LaFeAsO except that As was replaced by P,
exhibited paramagnetism in the normal conducting state[36]. This
prompted us that the ability of the P atoms in the Fe-P sandwich
structure to mediate the exchange interaction was rather weak. Based
on this fact, we conjecture that the appropriate doping in As
monolayers could modify the ratio $J_{2}/J_{1}$ value and possibly
to reach 1/2. It therefore deserves to explore the new material to
observe the expected phase transition.

In summary, we find that in a 2D AF system described by the
$J_{1}-J_{2}$ model, there can occur the phase transformation
between the collinear and N\'{e}el states under the condition that
$J_{2}/J_{1}=1/2$.


This work is supported by the National 973 Project 2012CB927402 and
the National Natural Science Foundation 11074145.


\begin{references}

\item[] {E-mail: wanghuaiyu@mail.tsinghua.edu.cn}

\item[${[1]}$] E. Daggotto and A. Moreo, Phys. Rev. Lett. \textbf{63}, 2148 (1989).
\item[${[2]}$] N. Tsyrulin \emph{et al}. Phys. Rev. B \textbf{81}, 134409 (2010).
\item[${[3]}$] R. Melzi \emph{et al}. Phys. Rev. Lett. \textbf{85}, 1318 (2000).
\item[${[4]}$] R. Melzi \emph{et al}. Phys. Rev. B \textbf{64}, 024409 (2001).
\item[${[5]}$] P. Carretta \emph{et al}. Phys. Rev. B \textbf{66}, 094420 (2002).
\item[${[6]}$] L. Bossoni \emph{et al}. Phys. Rev. B \textbf{83}, 014412 (2011).
\item[${[7]}$] Y. Kamihara \emph{et al}., J. Am. Chem. Soc. \textbf{130}, 3296 (2008).
\item[${[8]}$] H. H. Wen \emph{et al}., Europhysics Journal \textbf{82}, 17009 (2008).
\item[${[9]}$] H. Takahashi \emph{et al}., Nature \textbf{453}, 376 (2008).
\item[${[10]}$] C. D. L. Cruz \emph{et al}., Nature \textbf{453}, 899 (2008).
\item[${[11]}$] X. H. Chen \emph{et al}., Nature \textbf{453}, 761 (2008).
\item[${[12]}$] G. F. Chen \emph{et al}., Phys. Rev. Lett. \textbf{100}, 247002 (2008).
\item[${[13]}$] Z. A. Ren \emph{et al}., Europhys. Lett. \textbf{82}, 57002 (2008).
\item[${[14]}$] M. Rotter \emph{et al}. Phys. Rev. Lett. \textbf{101}, 107006 (2008).
\item[${[15]}$] J. E. Hirsch and S. Tang, Phys. Rev. B \textbf{39}, 2887 (1989);
O. P. Sushkov \emph{et al}., Phys. Rev. B \textbf{63}, 104420
(2001); H. U. Ueda and K. Totsuka, Phys. Rev. B \textbf{76}, 214428
(2007); R. Kumar and B. Kumar, Phys. Rev. B \textbf{77}, 144413
(2008); J. Richter \emph{et al}., Phys. Rev. B \textbf{81}, 174429
(2010); R. Shindou \emph{et al}., Phys. Rev. B \textbf{84}, 134414
(2011).
\item[${[16]}$] G. M. Zhang\emph{et al}., Phys. Rev. Lett. \textbf{91}, 067201
(2003); K. Takano \emph{et al}., Phys. Rev. Lett. \textbf{91},
067201 (2003).
\item[${[17]}$] P. Chandra and B. Doucot, Phys. Rev. B \textbf{38}, 9335 (1988);
H. A. Ceccatto \emph{et al}. Phys. Rev. B \textbf{45}, 7832 (1992);
K. Takano \emph{et al}.., Phys. Rev. Lett. 91, 197202 (2003); S.
Moukouri, Physics Letters A \textbf{352}, 256 (2006).

\item[${[18]}$] G. Misguich \emph{et al}., Phys. Rev. B \textbf{68}, 113409 (2003).
\item[${[19]}$] H. Rosner \emph{et al}., Phys. Rev. B \textbf{67}, 014416 (2003).
\item[${[20]}$] M. H\"{a}rtel \emph{et al}., Phys. Rev. B \textbf{81}, 174421 (2010).
\item[${[21]}$] N. Shannon\emph{et al}, Eur. Phys. J. B \textbf{38}, 599
(2004).

\item[${[22]}$] B Schmidt \emph{et al}., J. Phys.: Condens. Matter \textbf{19}, 145211 (2007).
\item[${[23]}$] B. Schmidt \emph{et al}., J. Magn. Magn. Mat. \textbf{310}, 1231 (2007).
\item[${[24]}$] J. R. Viana and J. R. D. Sousa, Phys. Rev. B \textbf{75}, 052403 (2007).
\item[${[25]}$] N. D. Mermin and H. Wagner, Phys. Rev. Lett. \textbf{17}, 1133 (1966).
\item[${[26]}$] B. Cheng \emph{et al}., Phys. Rev. B \textbf{85}, 144426 (2012).
\item[${[27]}$] F. Ma \emph{et al}., Phys. Rev. B \textbf{78}, 224517 (2008).
\item[${[28]}$] S. Ishibashi \emph{et al}., J. Phys. Soc. Japan \textbf{77}, 053709 (2008).
\item[${[29]}$] S. Ishibashi and K. J. Terakura, Phys. Soc. Japan \textbf{77}, Suppl. C, 91 (2008).
\item[${[30]}$] T. Yildirim, Phys. Rev. Lett. \textbf{101}, 057010 (2008).
\item[${[31]}$] E. B. Anderson and H. Callen, Phys. Rev. \textbf{136}, A1068 (1964).
\item[${[32]}$] P. Fr\"{o}brich and P. Kuntz, Physics Reports \textbf{432}, 223 (2006).
\item[${[33]}$] E. Manousakis \emph{et al}., Phys. Rev. B \textbf{78}, 205112 (2008).
\item[${[34]}$] Q. Huang \emph{et al}., Phys. Rev. B \textbf{78} 054529 (2008);
J. Zhao \emph{et al}., Nature Materials \textbf{7}, 953 (2008); J.
Zhao \emph{et al}., Phys. Rev. B \textbf{78}, 132504 (2008); Y. Chen
 \emph{et al}., Phys. Rev. B \textbf{78}, 064515 (2008); Y. Qiu \emph{et al}.,
Phys. Rev. Lett. \textbf{101}, 257002 (2008); M. Tropeano et al.,
Supercond. Sci. Technol. \textbf{21}, 095017 (2008).
\item[${[35]}$] A. I. Goldman \emph{et al}., Phys. Rev. B \textbf{78}, 100506R (2008); T. Yildirim, Phys. Rev. Lett. \textbf{102}, 037003 (2009);
 J. Zhao \emph{et al}., Phys. Rev. B \textbf{78}, 140405R (2008); J. Zhao \emph{et al}., Phys.
Rev. Lett. \textbf{101}, 167203 (2008); Q. Huang \emph{et al}.,
Phys. Rev. Lett. \textbf{101}, 257003 (2008).
\item[${[36]}$] Y. Kamihara \emph{et al}., Phys. Rev. B \textbf{77}, 214515 (2008).


\end{references}
\end{document}